\documentclass[aps,twocolumn,groupedaddress] {revtex4-1}
\usepackage{amsmath}
\usepackage{amsfonts}
\usepackage{amssymb}
\usepackage{graphicx}
\usepackage{caption}
\usepackage{subcaption}
\usepackage{float}
\usepackage{braket}
\allowdisplaybreaks
\newcommand {\bp}{\begin{pmatrix}}
\newcommand {\ep}{\end{pmatrix}}
\newcommand{\be}{\begin{equation}} \newcommand{\ee}{\end{equation}}
\newcommand{\bea}{\begin{eqnarray}}\newcommand{\eea}{\end{eqnarray}}

\DeclareMathOperator{\sech}{sech}
\DeclareMathOperator{\sgn}{sgn}
\begin{document}

\title{ Constructing Solvable Models of Vector Non-linear Schr$\ddot{o}$dinger Equation with
Balanced Loss and Gain via Non-unitary transformation}
\author{ Pijush K. Ghosh}
\email[]{pijushkanti.ghosh@visva-bharati.ac.in}
\affiliation{ Department of Physics, Siksha-Bhavana, Visva-Bharati University, 
Santiniketan, PIN 731 235, India.}
\date{\today}

\begin{abstract}
We consider vector Non-linear Schr$\ddot{o}$dinger Equation(NLSE) with balanced loss-gain(BLG),
linear coupling(LC) and a general form of cubic nonlinearity. We use a non-unitary transformation to
show that the system can be exactly mapped to the same equation without the BLG and LC,
and with a modified time-modulated nonlinear interaction. The nonlinear term
remains invariant, while BLG and LC are removed completely, for the special case of a pseudo-unitary
transformation. The mapping is generic and may be used to construct exactly solvable autonomous
as well as non-autonomous vector NLSE with BLG. We present an exactly solvable two-component
vector NLSE with BLG which exhibits power-oscillation. An example of a vector NLSE with BLG and arbitrary even number
of components is also presented.
\end{abstract}
\maketitle

The NLSE finds application in many diverse branches of modern science, including optics\cite{ac,sulem, soli-rev}, Bose-Einstein
Condensation(BEC)\cite{r2}, plasma physics\cite{r3}, gravity waves\cite{r4} and $\alpha$-helix protein dynamics\cite{r5}.
The homogeneous and autonomous NLSE with cubic nonlinearity is integrable in $1+1$ dimensions admitting soliton
solutions\cite{ac}. The NLSE has a rich mathematical structure\cite{sulem} and considered as one of the important examples in the
field of integrable  and exactly solvable models. Several generalizations of NLSE have been considered over
the years to describe and model various emerging physical phenomenon\cite{soli-rev,r2,mana,zs,ml,wang,dp,non-auto}. For example,
the study on homogeneous and autonomous NLSE paved the way for a better understanding of wave-propagation in non-linear
media in the context of optics\cite{soli-rev}. Similarly, investigations on inhomogeneous and non-autonomous
NLSE\cite{soli-rev,non-auto} became relevant after the experimental realizations of BEC\cite{r2}. With the emergence of
Parity-Time(${\cal{PT}}$) symmetric theory\cite{cmb} and its growing relevance in optics and other areas,  study on generalized
NLSE has been further diversified into several directions\cite{rev}. A few active areas of research in the context of ${\cal{PT}}$
symmetric theory are NLSE with ${\cal{PT}}$-symmetric confining complex potential\cite{complexp}, non-local
NLSE\cite{abl,ds-nlse}, NLSE with BLG\cite{rev,dm,fa,igor,pd,alex,dias}. 

The central focus of this article is on  NLSE with BLG in which the components of a vector NLSE are subjected to loss and gain
such that the net flow out of the system is zero,  i.e. the loss and gain are balanced.
A particular class of NLSE with constant\cite{dm} as well as time-dependent\cite{fa} BLG has been investigated earlier.
In the terminology of optics, the nonlinear interaction contains both self-phase modulation as well as cross-phase modulation terms.
The system has been investigated from the viewpoint of solitons, modulational instability, ${\cal{PT}}$-symmetry breaking,
exceptional points etc. and the relevant results are nicely reviewed in Ref. \cite{rev}. All these investigations are mostly based on
numerical and approximate methods. Integrable and/or exactly solvable NLSE with BLG are still elusive. Within this background,
we investigate a more general class of NLSE with BLG which contains terms related to four-wave mixing in addition to self-phase
and cross-phase modulation terms. We present a generic method to investigate such systems analytically leading to exact solutions
under certain conditions.

We show that the BLG and LC terms may be removed completely from a class of NLSE with BLG via a non-unitary transformation
with its effect manifested in the time-dependence of the nonlinear term. The non-unitary transformation may be viewed  as a gauge
transformation involving complex scalar gauge potential. A real scalar potential corresponds to the vanishing loss-gain terms
and the two systems become gauge-equivalent, since the non-unitary transformation reduces to a unitary transformation in this
limit. The power-spectra and other observables have same time-dependence for both the cases. However, for a non-vanishing imaginary
part of the complex scalar potential, the original and the mapped systems are not gauge equivalent and the observables have different
time-dependence. We show that exact and analytical solutions of a class of non-autonomous vector NLSE with specified time-dependence
may be found by mapping it to solvable autonomous system via non-unitary transformations. The time-dependence of observables of these
two systems are different due to the non-unitary nature of the transformation that connects them. There is a special class
of non-unitary transformation, namely  pseudo-unitary\cite{ali} transformation, for which 
the NLSE with BLG can be mapped to the same NLSE without the loss-gain terms. The nonlinear term remains invariant under 
pseudo-unitary mapping, thereby, a given (non-)autonomous system is mapped to (non-)autonomous system. This allows to construct
exactly solvable models of NLSE with BLG by mapping them to known solvable models of NLSE. We present a few examples of NLSE with
BLG in detail to exemplify the general result.

The vector NLSE is introduced in terms of a
$N$-component complex scalar field $\Psi$ and its hermitian adjoint $\Psi^{\dagger}$ as,
\bea
i \left (I \frac{\partial}{\partial t}+i A \right)\Psi= 
- \frac{\partial^2 \Psi}{\partial x^2} -\delta \left ( \Psi^{\dagger} M \Psi \right )\Psi
\label{nlse}
\eea
\noindent where $I$ is the $N \times N$ identity matrix and $\delta$ is a real parameter. The $N \times N$
non-hermitian matrix $A$ is decomposed in terms of two hermitian matrices $B$ and $C$
as $A=B + i C$ with the condition that $C$ is a traceless diagonal matrix.
The loss-gain terms in Eq. (\ref{nlse}) are described by the term $-i C \Psi$,
while the LC among different field-components are governed by
$B \Psi$. The $N \times N$ hermitian and non-singular matrix $M$ does not depend on
complex scalar fields and Eq. (\ref{nlse}) describes a coupled cubic nonlinear
Schr$\ddot{o}$dinger equation with BLG. The space-time modulation
of the nonlinear strengths may be incorporated via explicit space-time dependence of $M$.

The NLSE in Eq. (\ref{nlse}) may be obtained from the Lagrangian density, 
\bea
{\cal{L}} & = &
\frac{i}{2} \left [ \Psi^{\dagger} M \left ( D_0 \Psi\right ) - 
(D_o \Psi)^{\dagger} M \Psi \right ]\nonumber \\
& - & \frac{\partial \Psi^{\dagger}}{\partial x}
M\frac{\partial \Psi}{\partial x}
+  \frac{\delta}{2} \left ( \Psi^{\dagger} M \Psi \right )^2
+\Psi^{\dagger} F_1 \Psi
\label{lag}
\eea
\noindent where the operator $D_0:=I \frac{\partial}{\partial t} + i A $ has formal
resemblance with the temporal component of covariant derivative with non-hermitian gauge potential
$A$ and the anti-hermitian matrix $F_1:=\frac{1}{2} \left ( A^{\dagger}M-M A \right )$. The hermitian
adjoint of Eq.  (\ref{nlse}) does not describe the equation satisfied by $\Psi^{\dagger}$ for $F_1
\neq 0$, rather it describes the equation obeyed by $\Psi^{\dagger}$ of a system whose
Lagrangian density is ${\cal{L}}^{*}$, i.e. complex conjugate of ${\cal{L}}$. This is because
${\cal{L}}$ is complex for $F_1 \neq 0$. The equation satisfied by $\Psi^{\dagger}$ has to be
derived directly by using the Euler-Lagrange equation for ${\cal{L}}$.
The conjugate momenta corresponding to $\Psi$ and $\Psi^{\dagger}$ are $\Pi_{\Psi}=
\frac{i}{2} \Psi^{\dagger} M$ and $\Pi_{\Psi^{\dagger}}=- \frac{i}{2} M \Psi$, respectively.
The Hamiltonian density ${\cal{H}}$ corresponding to ${\cal{L}}$ has the form,
\bea 
{\cal{H}} = \frac{\partial \Psi^{\dagger}}{\partial x} M \frac{\partial \Psi}{\partial x}
-\frac{\delta}{2} \left ( \Psi^{\dagger} M \Psi \right )^2 + \Psi^{\dagger} M A \Psi\nonumber
\label{hami}
\eea
\noindent The effect of the BLG is contained in the mass term
$\Psi^{\dagger} M A \Psi$, which in general is complex-valued and
becomes real-valued only for $F_1=0$, i.e. for a $M$-pseudo-hermitian $A$. 
The Hamiltonian is real-valued for the same condition, since the first
two terms are real-valued irrespective of a pseudo-hermitian $A$.
Consequently, a quantized Hamiltonian $\int dx {\cal{H}}$ is non-hermitian
without the pseudo-hermiticity condition and is expected to be hermitian
for $F_1=0$ with suitable quantization condition. In general, the energy
$E=\int dx dt {\cal{H}}$ may not have a lower bound leading to collapse
of the wave-function $\Psi$. However, $E \geq 0$ for a $M$-pseudo-hermitian
$A$ with a positive-definite $M$ and semi-positive-definite $A$.

We use a non-unitary transformation relating
$\Psi$ with a $N$-component complex scalar field $\Phi$ as follows,
\bea
\Psi(t,x) = U(t) \Phi(t,x), \ \ U(t)=e^{-i A t}
\label{non-uni}
\eea
\noindent which when substituted in Eq. (\ref{nlse}) results in the equation,
\bea 
&& i \frac{\partial \Phi}{\partial t} =  - \frac{\partial^2 \Phi}{\partial x^2} -\delta
\left ( \Phi^{\dagger} G \Phi \right ) \Phi,
G= U^{\dagger} M U
\label{m-nlse}
\eea
\noindent The time-dependent non-unitary transformation removes the
loss-gain and the LC terms by modifying the nonlinear interaction.
The nonlinear term remains unchanged
due to the transformation if and only if $G=M$, i.e. $U$ is pseudo-unitary\cite{ali}
with respect to $M$ or equivalently $A$ is $M$-pseudo-hermitian,
\bea
U^{\dagger} M U =  M  \Leftrightarrow A^{\dagger}=M A M^{-1}
\eea
The pseudo-hermiticity of $A$ can also be derived by expanding $G(t)$ in powers of $t$ 
with the identification of $F_0=M$,
\bea
G(t)=\sum_{n=0}^{\infty} \frac{(i t)^n}{n!} F_n, F_{n+1}=A^{\dagger} F_n- F_n A\nonumber
\eea 
\noindent The condition $G=M$ leads to $F_1=0$, i.e. $A$ is $M$-pseudo-hermitian.
{\em The first important result is that Eq. (1) with $M$-pseudo-hermitian $A$  can
be mapped to the same equation without the loss-gain and the LC terms as given
in Eq.(\ref{m-nlse}).}
Further, if the transformed equation (\ref{m-nlse}) with $G=M$ is exactly solvable, solutions for
Eq. (\ref{nlse}) can be obtained by using the pseudo-unitary transformation.

The second important result concerns the case for which $A$ is neither hermitian
nor $M$-pseudo-hermitian or equivalently $U$ is neither unitary nor pseudo-unitary.
The matrix $M$ for a non-unitary $U$ may be chosen to be time-dependent such that Eq. (\ref{nlse}) is
necessarily non-autonomous, while Eq. (\ref{m-nlse}) is autonomous. We choose the matrix $M$ in
terms of real parameters $\alpha_j$ as,
\bea
M(t)= \sum_{j=0}^{N^2-1} \alpha_j \left [ U^{\dagger}(- t) \lambda_j U(-t) \right ]
\label{expand-m}
\eea
\noindent where the constant matrices $\lambda_j$ denote a suitable basis for expanding $M$ and $G$
with $\lambda_0$ being the identity matrix.
The matrix $G$ for the choice of $M$ in Eq. (\ref{expand-m}) has the form $G=\sum_j \alpha_j
\lambda_j$ and Eq. (\ref{m-nlse}) reduces to integrable Manakov system\cite{mana} of coupled
vector NLSE for $G= \alpha_0 \lambda_0$, which may be realized by choosing all $\alpha_j=0$ except
$\alpha_0$.  The solution of the non-autonomous equation (\ref{nlse}) may be found from the
solution of Eq. (\ref{m-nlse}) by using the
non-unitary transformation. Various integrable and/or solvable generalizations of Manakov
systems are known\cite{zs,ml,wang,dp}. The parameters $\alpha_j$ may be chosen appropriately
to find solvable non-autonomous system with BLG and LC corresponding to these
known solvable models. There is an useful duality relation between $M$ and $G$. The matrix $M(t)$
in Eq. (\ref{expand-m}) is time-dependent for a constant $G$. If $M$ is chosen as time-independent
$M=\sum_j \alpha_j \lambda_j$, then $G$ becomes time-dependent $G(t)=M(-t)$ where $M(t)$ is
given by Eq. (\ref{expand-m}).

The transformation in Eq. (\ref{non-uni}) may be used to remove the loss-gain
terms completely even for $Tr(C) \neq 0$, i.e. the case of unbalanced loss-gain.
However, the non-unitary matrix $U$ and hence, $\Psi$ necessarily contains a term
growing/decaying in time for $Tr(C) \neq 0$. In general, the $N \times N$ matrices $B$ and
$C$ may be expressed as generators of $SU(N)$ in the fundamental representation as, 
\bea
B=\frac{1}{2} \sum_{\substack{i=1\\(i\neq j^2-1)}}^{N^2-1} \beta_i \lambda_i,
C=\frac{1}{2} \sum_{i=2}^N \beta_{i^2-1} \lambda_{i^2-1}, \ j=2, \dots, N,\nonumber
\eea
\noindent where $Tr(\lambda_i \lambda_j)=2 \delta_{ij}$, $\lambda_{i^2-1}$ are diagonal and 
consequently, $Tr(B)=Tr(C)=0$. We now replace
$A$ by $\tilde{A}=B + i ( C+\beta_0 I), \beta_0 \in \mathbb{R}$, where $I$ is the
$N \times N$ identity matrix. The loss-gain is now unbalanced for $\beta_0 \neq 0$, since 
$Tr(\tilde{C})= Tr(C+\beta_0 I)= N \beta_0$.
The non-unitary matrix $\tilde{U}:=e^{-i \tilde{A}t}=e^{\beta_0 t} U$ and $U$ may be
expressed in terms of $N$ eigenvalues $e_j$ of $A$ as\cite{dimitri},
\bea
{U}=I \frac{1}{N} K - i \sum_{j=1}^{N^2-1} \lambda_j
\frac{\partial K}{\partial (t \beta_j)}, \
K(\beta,t)=\sum_{j=1}^N e^{i e_j t}\nonumber
\eea
\noindent with the condition $\sum_{j=1}^N e_j=0$, since $Tr(A)=0$. The eigenvalues
$e_j$ are functions of the parameters $\beta_j$ and $U$ is periodic
in time in a region in the parameter space for which all $e_j$'s are real. The condition
$\sum_{j=1}^N e_j=0$ implies that a common factor of the form $e^{-\beta_0 t}$
can not be taken out of the matrix $U$ to cancel the multiplicative
term $e^{\beta_0 t}$ appearing in $\tilde{U}$ and still making $\tilde{U}$ periodic in time.
This leads to unbounded/decaying solution $\Psi$ for any periodic or soliton
solution $\Phi$ of Eq. (\ref{m-nlse}).
Thus, an unbalanced gain-loss in the system
leads to growing and decaying solutions for $\beta_0 > 0$ and $\beta_0 <0$, respectively.
This is the reason for restricting the discussions to the case of balanced loss-gain only.

There are previous studies\cite{rev,dm,igor,pd,alex,dias,fr1,PGK,fr2,fr3,fr4,fr5,fr6}
to remove  BLG and/or the LC terms from Eq. (\ref{nlse}) through appropriate transformations
and under certain reductions of the original equation. The transformations for all these
cases are necessarily unitary, while the transformation used in this article
is non-unitary. This is a major difference \textemdash systems related by unitary
transformation are gauge equivalent, while the same can not be claimed for systems
related by non-unitary transformation. This is manifested in the result that the power
of the standard Manakov system $P_{\Phi}=\Phi^{\dagger} \Phi$ is different from the
power $P_{\Psi}=\Psi^{\dagger} \Psi$ of Manakov system with BLG, although
they are connected via a non-unitary transformation. In particular,
$P_{\Psi}=\Psi^{\dagger} \Psi=\Phi^{\dagger} \left ( U^{\dagger} U \right ) \Phi
\neq P_{\Phi}$ and $P_{\Psi}$ will be calculated explicitly to highlight this feature
for the examples considered in this article. Similarly, one can show that the time-dependence
of other observables like square of the width of the wave-packet $I_1=\int dx x^2 P_{\Psi}$
and its speed of growth $I_2= - i \int dx \left ( \Psi^{\dagger} \frac{\partial \Psi}{\partial
x} - \frac{\partial \Psi^{\dagger}}{\partial x} \Psi \right )$ are different for systems
connected via non-unitary or pseudo-unitary transformation. Unitary transformations have been
used in physics in different contexts, particularly in the context of field theory, for past several
decades.  It seems that the pseudo-unitary invariance of a Hamiltonian system and its use to construct
exact solution have not been considered earlier. Further, the non-unitary mapping to remove the BLG
 and the LC terms is exact and unlike the previous investigations\cite{rev,dm,igor,pd,alex,dias,PGK},
no reduction of the original equation is considered. Thus, the mapping proposed in this article is new
compared to the existing methods to remove BLG and/or LC terms via appropriate transformations.

We present an example of a two-component NLSE to elucidate the general results by choosing,
\bea
G= \sum_{j=0}^3 \alpha_j \sigma_j
\label{g1}
\eea
\noindent where $\sigma_0$ is the $2 \times 2$ identity matrix and
$\sigma_1, \sigma_2, \sigma_3$ denote the Pauli matrices with $\sigma_3$ being diagonal.
The terms in $\Phi^{\dagger} G \Phi$ in Eq. (\ref{m-nlse}) has standard
physical interpretation. In particular, in the terminology of optics, terms containing $\alpha_0$ and $\alpha_3$
are related to self-phase and cross-phase modulations, while terms which
include $\alpha_1, \alpha_2$ describe the effect of  four-wave mixing.
Eq. (\ref{m-nlse}) for the above choice of $G$ is integrable for any values of the
real parameters $\alpha_j$\cite{wang}.  The celebrated
Manakov system\cite{mana} of two coupled NLSE is obtained for $\alpha_3=0, \alpha
\equiv \alpha_1+i\alpha_2=0$, while $\alpha_0=0, \alpha=0$ correspond to Zakharov-Schulman system\cite{zs}.

The non-hermitian matrix $A$ in Eq. (\ref{nlse}) is chosen as,
\bea
A=\beta_1 \sigma_1 + \beta_2 \sigma_2 + i \Gamma \sigma_3
\label{non-hami}
\eea
The real constants $\beta_{1,2}$ linearly couple
two components of $\Psi$, while $\Gamma$ measures the loss-gain strength.
The matrix $A$ is $M$-pseudo-hermitian for the conditions,
\bea
\alpha_3=0, \ \ \frac{\alpha_0}{\vert \alpha \vert} = \frac{\vert \beta \vert}{\Gamma} \sin(\theta_{\alpha}-\theta_{\beta})
\label{condi}
\eea
\noindent for which $G=M$ is time-independent, where $\alpha ={\vert \alpha \vert} e^{i \theta_{\alpha}}$ and
$\beta \equiv \beta_1 + i \beta_2={\vert \beta \vert} e^{i \theta_{\beta}}$. The matrix $M$ with $\alpha_3=0$ is
positive definite for $\alpha_0 > {\vert \alpha \vert}$ and the second condition of Eq. (\ref{condi}) is
consistent for the choice ${\vert \beta \vert} > {\vert \Gamma \vert}$ with $0 < \theta_{\alpha}-
\theta_{\beta} < \pi$ for $\Gamma > 0$ and $\pi < \theta_{\alpha}-\theta_{\beta} <2 \pi$
for $\Gamma < 0$. The matrix $M$ can be rewritten after imposing the condition
of pseudo-hermiticity as,
\bea
M= \frac{{\vert \alpha \vert} {\vert \beta \vert}}{\Gamma} \sin(\theta_{\alpha}-\theta_{\beta}) \sigma_0 + \alpha_1
\sigma_1 + \alpha_2 \sigma_2
\label{auto}
\eea
\noindent The NLSE in Eq. (\ref{nlse}) with $\delta=1$ is solvable for $A$ and $M$ given by Eqs. (\ref{non-hami}) and 
(\ref{auto}), respectively. The non-unitary operator $U$ connecting $\Psi$ and $\Phi$ may be expressed
as a $2 \times 2$ matrix in terms of the parameter $\theta \equiv \sqrt{{\vert \beta \vert}^2-\Gamma^2}$,
\bea
U = \sigma_0 \cos (\theta t) - \frac{i A}{\theta} \sin(\theta t) 
\ \textrm{for} \ \theta \neq 0
\label{uexp}
\eea
\noindent The parameter $\theta$ becomes purely imaginary
for $\Gamma^2 > {\vert \beta \vert}^2$ and the periodic functions change to the
corresponding hyperbolic functions. Consequently, a bounded solution for $\Phi$ will correspond
to an unbounded solution for $\Psi$ in the long time behaviour. This is true for $\theta=0$ also
for which $U$ has a linear time-dependence. The loss-gain parameter is restricted within the
range $-{\vert \beta \vert} < \Gamma < {\vert \beta \vert}$ so that time-dependence
of $U$ is periodic. It may be noted that this is also one of the consistency conditions for
$A$ to be $M$-pseudo-hermitian with a positive-definite $M$.

\begin{widetext}

\begin{figure}[ht!]
\begin{subfigure}{.33\textwidth}
\centering
\includegraphics[width=.8\linewidth]{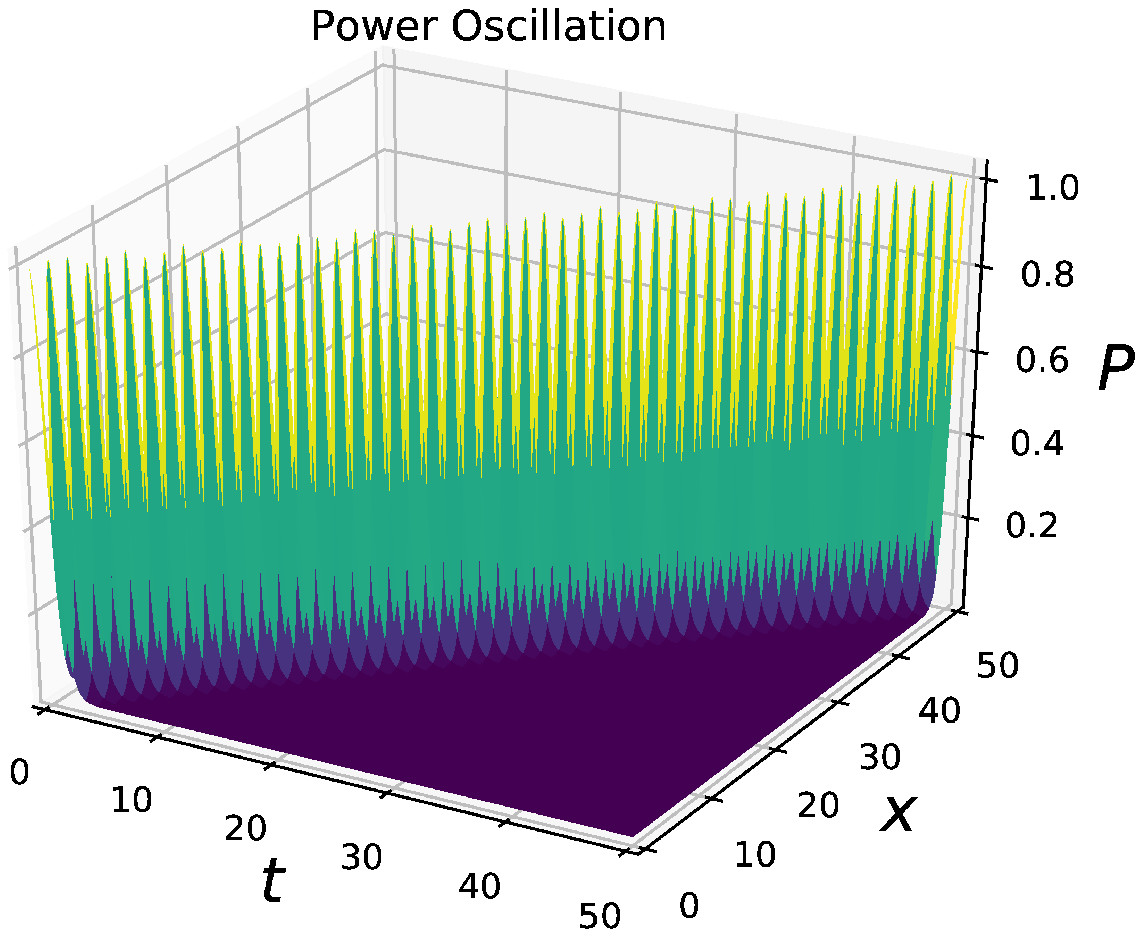}
\caption{ $\Gamma=0.1$}
\label{pp1}
\end{subfigure}%
\begin{subfigure}{.33\textwidth}
\centering
\includegraphics[width=.8\linewidth]{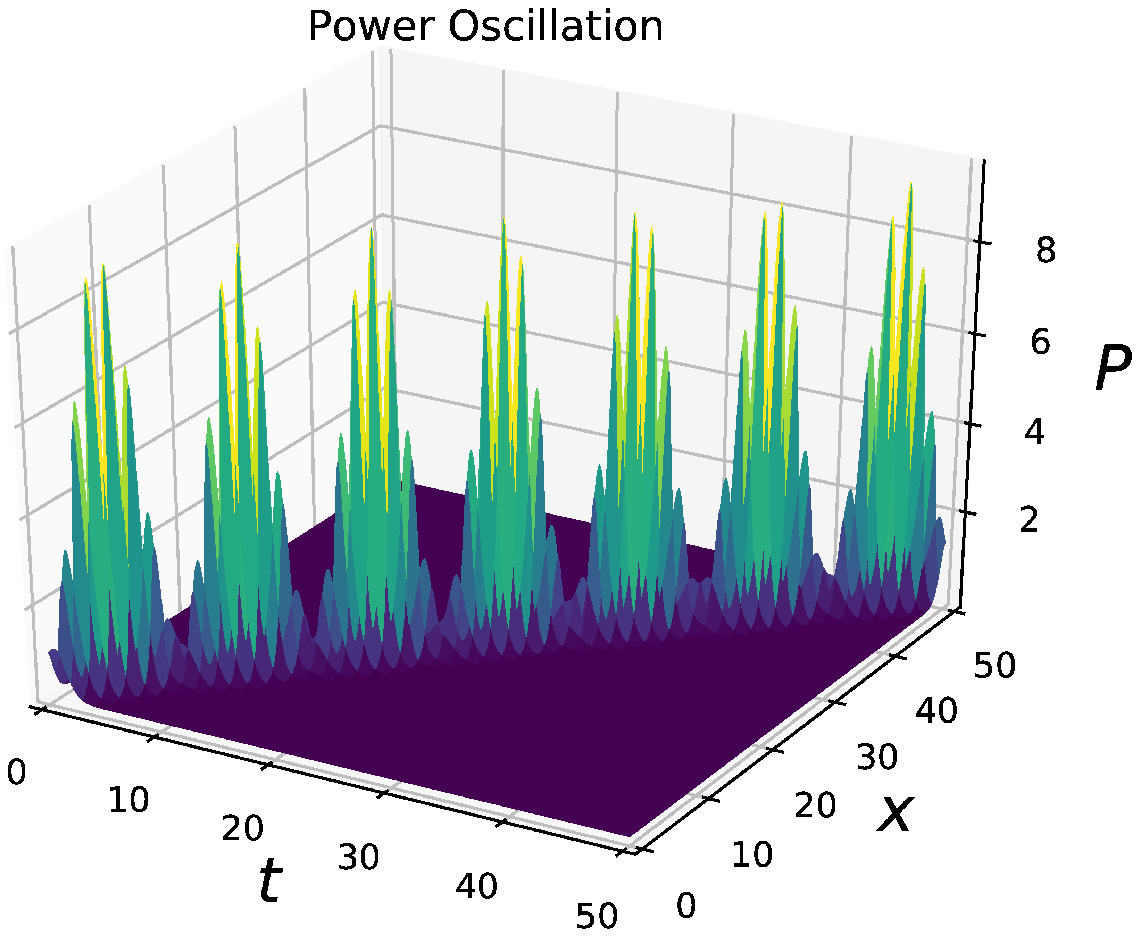}
\caption{ $\Gamma=0.9$}
\label{pp1}
\end{subfigure}%
\begin{subfigure}{.33\textwidth}
\centering
\includegraphics[width=.8\linewidth]{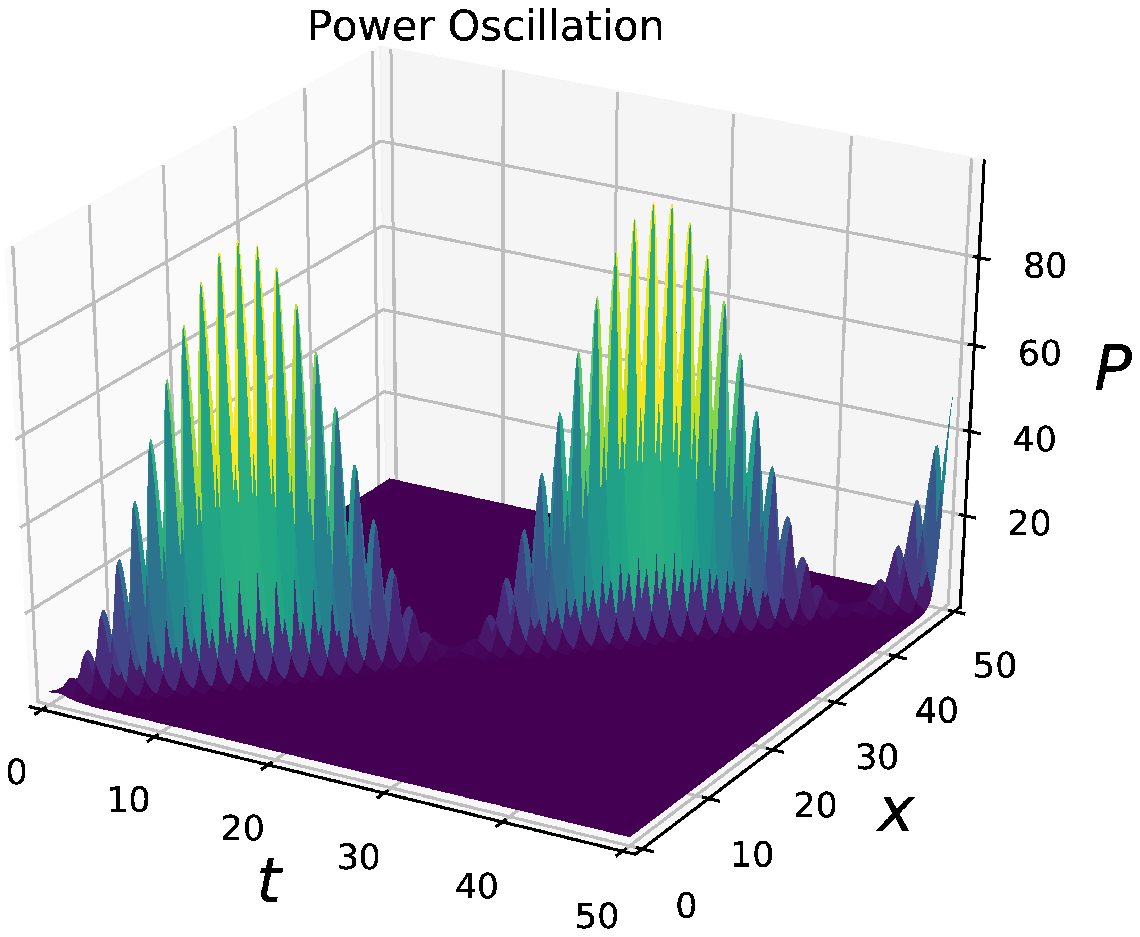}

\caption{$\Gamma=.99$}
\label{pp2}
\end{subfigure}%
\caption{(Color online) Plots of $P(x,t)$  for $\kappa=v=\beta=1$ and three different
values of $\Gamma=0.1, 0.9, 0.99$}
\label{power-oscii-1}
\end{figure}

\end{widetext}

The solution of Eq. (\ref{m-nlse}) for $\delta=1$ and $G=M$ given by Eq. (\ref{auto}) has the expression\cite{wang},
\bea
\Phi=\sqrt{\frac{2}{C}} \ \kappa W \sech \left [\kappa (x-vt) \right ]e^{i(\frac{vx}{2}-\omega t)}
\label{sol1}
\eea
\noindent where $\omega=\frac{v^2}{4} -\kappa^2$, $C=W^{\dagger}M W$
and $W$ is an arbitrary two-component constant complex vector. The semi-positivity of $M$ ensures that the constant
$C$ is semi-positive. The constants $v, \omega, \kappa$ correspond to the velocity, propagation constant and amplitude,
respectively for the one soliton solution $\Phi$. The power $P_{\Psi}$ for the loss-gain system oscillates with time,
\bea
&& P_{\Psi}= \frac{2 \kappa^2 W^{\dagger} W}{\vert C  \vert} \sech^2\left [\kappa
\left ( x-v t \right ) \right ] N(t)\nonumber \\
&& N(t) = 1 + N_1 \sin^2(\theta t) + N_2 \sin(2 \theta t)
\label{power1}
\eea
\noindent  where $N_1= \frac{2 \Gamma}{\theta^2} (\Gamma + \frac{ -\beta_2 C_1 +
\beta_1 C_2}{C_0} )$, $N_2= \frac{\Gamma C_3}{\theta C_0}$ and $C_j=W^{\dagger} \sigma_j W, j=0,1,2,3$.
The power-oscillation vanishes for no loss-gain in the system, i.e. $\Gamma=0$. The condition $N(t) \geq 0 \ \forall
\ t$ may be implemented in several ways by choosing the integration constants and parameters appropriately. For example,
the constant $N_2$ vanishes if the two components of the complex
vector $W$ are chosen as $W_0 e^{i \theta_{W_j}}, j=1,2$ such that they differ only in phases. Further,
fixing $\theta_{W_1}=\theta_{W_2} +\theta_{\beta} + (n+1) \pi, n \in \mathbb{Z}$, the constant
$N_1=\frac{2 \Gamma^2}{\theta^2}$ becomes semi-positive definite. The power oscillation may be visualized in the
plots of $P(x,t)=\frac{ {\vert C \vert}}{2 \kappa^2 W^{\dagger} W} P_{\Psi}$ in Fig. \ref{power-oscii-1}
with the above choices of constants and parameters for $\kappa=v={\vert \beta \vert}=1$ and three values of
$\Gamma=0.1, 0.9, 0.99$. The amplitude and time-period of oscillation grows as $\Gamma$ is increased and approaches
$\beta$. The peak of the power-oscillation amplifies approximately by 10 times as ${\vert \beta \vert}$ is increased by $.09$.
The solution becomes unbounded for $\Gamma \geq {\vert \beta \vert}$. The loss-gain parameter $\Gamma$ may be used as
a controlling parameter for power-oscillation. 

\begin{widetext}

\begin{figure}[ht!]
\begin{subfigure}{.33\textwidth}
\centering
\includegraphics[width=.8\linewidth]{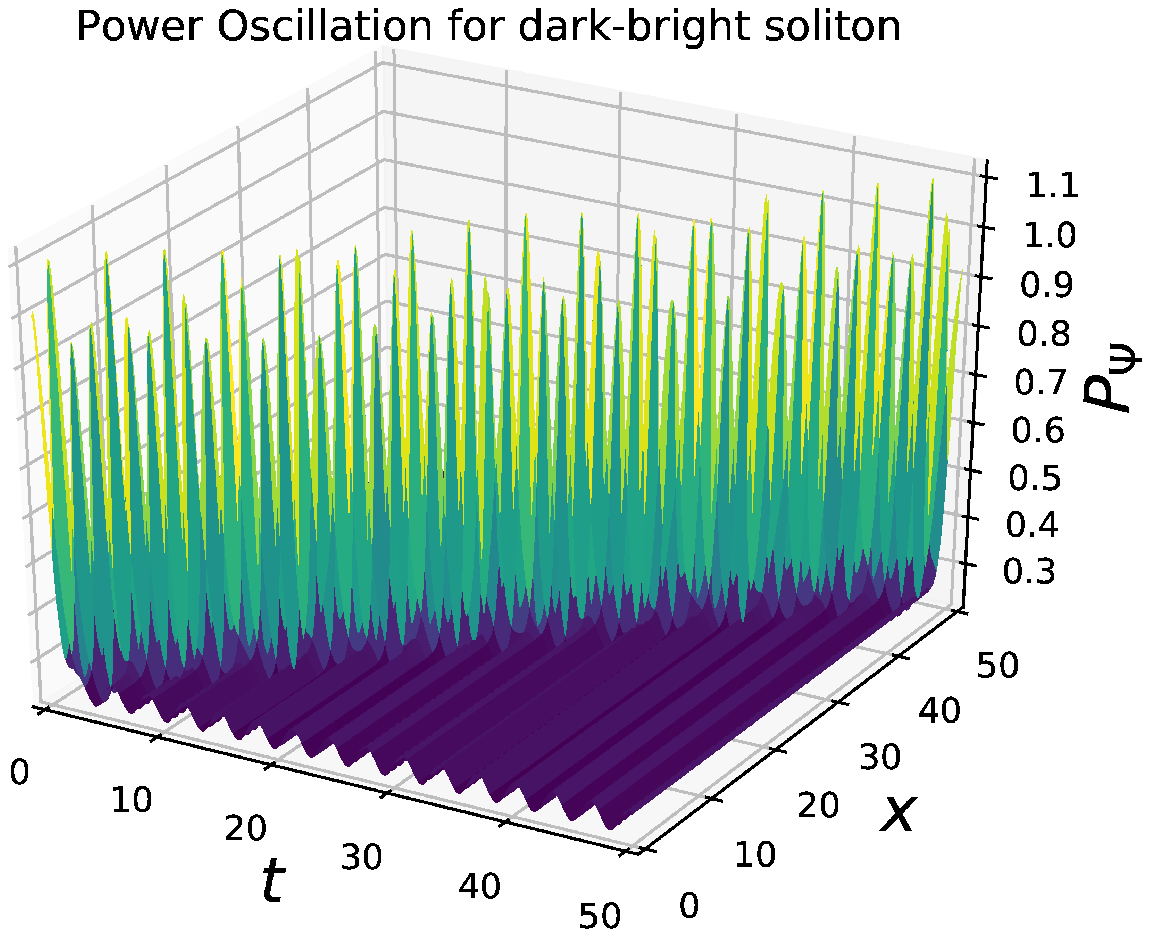}
\caption{$\Gamma=0.1$}
\label{pp3}
\end{subfigure}%
\begin{subfigure}{.33\textwidth}
\centering
\includegraphics[width=.8\linewidth]{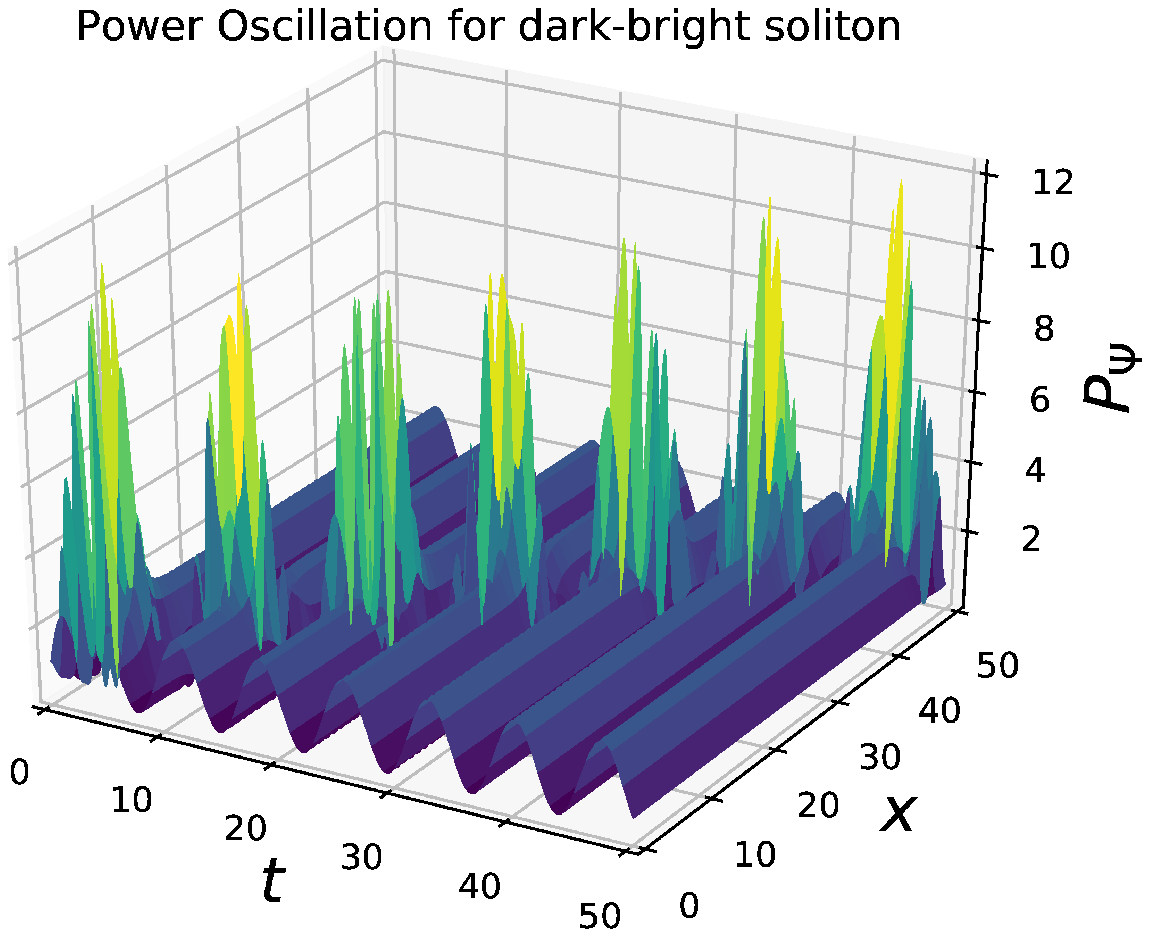}
\caption{$\Gamma=0.9$}
\label{pp3}
\end{subfigure}%
\begin{subfigure}{.33\textwidth}
\centering
\includegraphics[width=.8\linewidth]{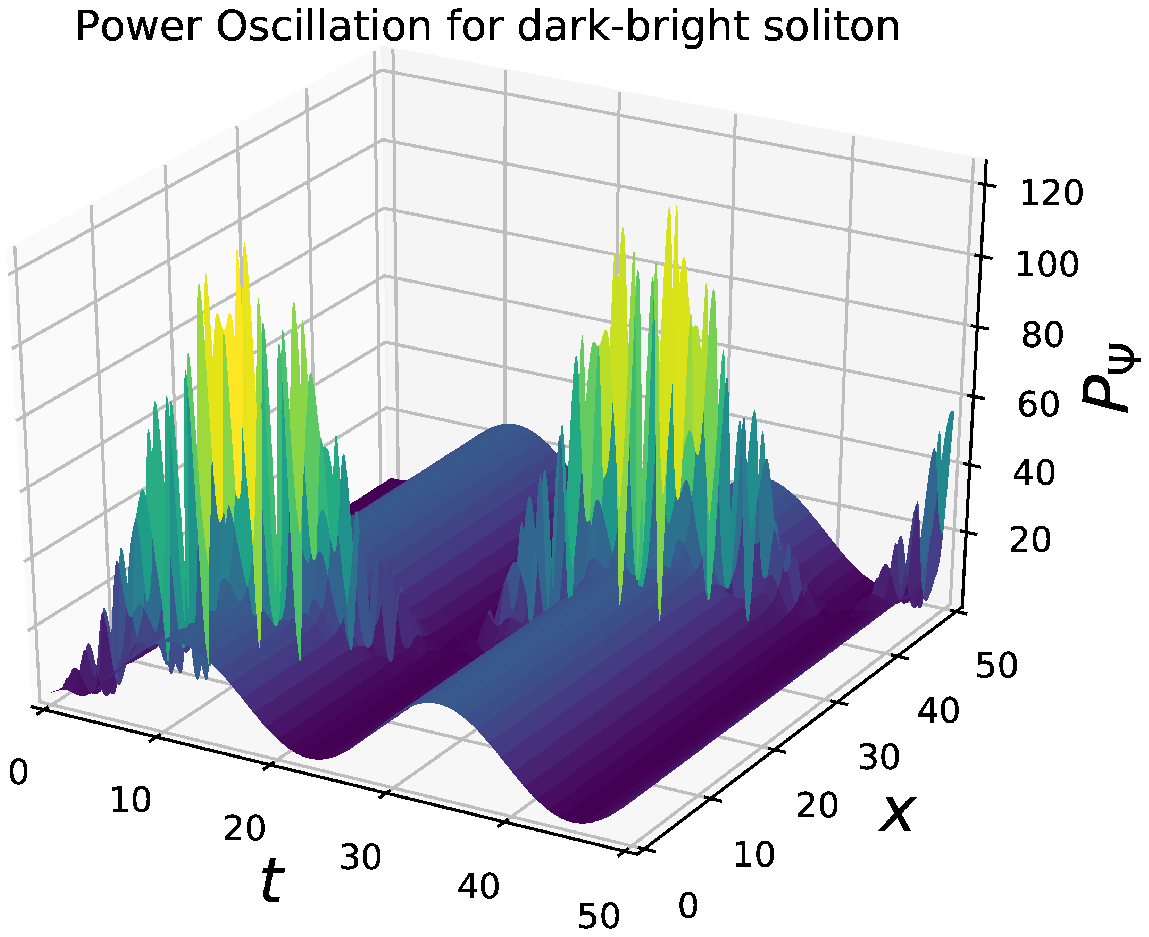}
\caption{$\Gamma=0.99$}
\label{pp4}
\end{subfigure}%
\caption{(Color online) Plot of $P_{\Psi}(x,t)$  for $a=v=\beta=1$, $b=.5$,
$\theta_{\beta}=0$ and three different values of $\Gamma=0.1, 0.9, 0.99$}
\label{power-oscii-2}
\end{figure}

\end{widetext}

We now discuss the situation where $A$ is not $M$-pseudo-hermitian,
and consequently, there is no restriction on the choice of the parameters $\alpha_i$'s as given in Eq.
(\ref{condi}) and $U(t)$ is not pseudo-unitary. The matrix $M(t)$ corresponding to $G$ and $A$
given in Eqs. (\ref{g1}) and (\ref{non-hami}), respectively, may be evaluated by using
Eq. (\ref{expand-m}) and substituting $\lambda_j=\sigma_j$. This leads to the expression,
\begin{widetext}
\bea
M(t) & = & \sigma_0 \left [ \alpha_0 + \frac{2 \Gamma \eta}{\theta^2} \sin^2(\theta t)-
\frac{\Gamma \alpha_3}{\theta} \sin(2 \theta t) \right ] +
\sigma_1 \left [ \alpha_1 - \frac{2 \beta_2 \eta}{\theta^2} \sin^2(\theta t) +
\frac{\beta_2 \alpha_3}{\theta} \sin(2 \theta t) \right ]\nonumber \\
& + & \sigma_2 \left [ \alpha_2 + \frac{2 \beta_1 \eta}{\theta^2} \sin^2(\theta t)
-\frac{\beta_1 \alpha_3}{\theta} \sin(2 \theta t) \right ] +
\sigma_3 \left [  \alpha_3 \cos(2 \theta t) -\frac{\eta}{\theta} \sin(2 \theta t) \right ],
\label{mexp}
\eea
\end{widetext}
\noindent where $\eta\equiv \Gamma \alpha_0-{\vert \beta \vert} {\vert \alpha \vert}
\sin(\theta_{\alpha} -\theta_{\beta})$
and in the limit of $M$-pseudo-hermitian $A$, i.e. Eq. (\ref{condi}) holds, the result $M=G$ is
reproduced. The NLSE in Eq. (\ref{nlse}) with the above $M(t)$ and $A$ in Eq. (\ref{non-hami}) is
mapped via the non-unitary transformation to the  equation,
\bea 
&& i \frac{\partial \Phi}{\partial t} =  - \frac{\partial^2 \Phi}{\partial x^2} -\delta
\sum_{j=0}^3 \alpha_j \left ( \Phi^{\dagger} \sigma_j \Phi \right ) \Phi,
\label{npsu-nlse}
\eea
\noindent which admits various exact analytical solutions. For the generic values of the parameters
$\alpha_i$, the expressions for $\Psi$ and $P_{\Psi}$ are given by Eqs. (\ref{sol1}) and (\ref{power1}),
respectively, where the factor $C$ in $\Psi, P_{\Psi}$ is evaluated without the condition (\ref{condi}).
The constant $C$ is an overall multiplication factor and does not affect the physical behaviour of the system
and can be chosen to be positive-definite for $\alpha_0 \geq \sqrt{{\vert \alpha \vert}^2 + \alpha_3^2 }$.
We present another solution by choosing $\delta=2, \alpha_3=1, \alpha_0=\alpha_1=\alpha_2=0$
for which $M(t)$ takes the form,
\bea
M=
\begin{pmatrix}
\cos(2 \theta t) - \frac{\Gamma}{\theta} \sin(2 \theta t) && i \frac{\beta^*}{\theta}
\sin(2 \theta t)\\
-i \frac{\beta}{\theta} \sin(2 \theta t) && -\cos(2 \theta t)- \frac{\Gamma}{\theta} \sin(2 \theta t)\nonumber
\end{pmatrix}
\label{mspecial}
\eea
\noindent 
and Eq. (\ref{npsu-nlse})
admits several exact solutions\cite{kanna}. We consider the bright-dark one soliton solution,
\bea
&& \Phi_1= a e^{i \left (\frac{vx}{2} -\omega_1 t \right )} \sech \left [ \kappa \left (x-v t\right) \right ]\nonumber \\
&& \Phi_2= b e^{i \left (\frac{vx}{2} -\omega_2 t \right )} \tanh \left [ \kappa \left (x-v t\right)\right]
\label{phi-special}
\eea
\noindent where $\kappa^2=a^2+b^2$, $\omega_1=\frac{v^2}{4} + b^2-a^2$ and $\omega_2=\frac{v^2}{4}+2 b^2$.
The solution $\Psi$ of Eq. (\ref{nlse}) for $M(t)$ given above
is obtained as $\Psi=U \Phi$,
where $U$ and $\Phi$ are given by Eqs. (\ref{uexp}) and (\ref{phi-special}), respectively. The power has the
expression:
\begin{widetext}
\bea
&& P_{\Psi}(x,t)= \left [ 1 + \frac{2 \Gamma^2}{\theta^2} \sin^2(\theta t) \right ]
\left \{ a^2  \sech^2 \left [ \kappa \left (x-v t\right)\right] + b^2 \tanh^2
\left [ \kappa \left (x-v t\right) \right ]\right \}
+ \frac{\Gamma}{\theta} \sin(2 \theta t) \left \{ a^2 - \kappa^2 \tanh^2 \left [
\kappa \left (x-vt \right ) \right ] \right \}\nonumber \\
&& - \frac{4 a b {\vert \beta \vert} \Gamma}{\theta^2} \sin^2(\theta t) \sin \left (\theta_{\beta}+\kappa^2 t\right)
\left \{ \sech \left [ \kappa \left (x-v t\right) \right ] \tanh \left [ \kappa \left (x-v t\right)\right]\right \}
\eea
\end{widetext}
\noindent which is plotted in Fig. \ref{power-oscii-2} for $a=v=\beta=1$, $b=.5$ and three different
values of $\Gamma=0.1, 0.9, 0.99$. The amplitude and time-period grows as $\Gamma$ approaches $\beta$ and the solution becomes
unbounded for $\Gamma \geq \beta$. There are other solutions of Eq. $(\ref{phi-special})$ leading to the same 
qualitative behaviour for $P_{\Psi}$ which will not be pursed in this article.

The results can be generalized easily to multi-component vector NLSE with appropriate choice
of the matrices $A$ and $M$ for $N >2$. We denote $A$ and $M$ appearing in Eq. (\ref{nlse}) for $N=2m$ as
$A_{2m}$ and $M_{2m}$, respectively.  The matrix $A_{2m}$ is chosen as $A_{2m}= I_m \otimes A$, 
where $I_{m}$ is the $m \times m$ identity matrix and the $2 \times 2$ matrix $A$
is given in Eq. (\ref{non-hami}).  The NLSE in Eq. (\ref{nlse}) with
this specific $A_{2m}$ describes pair-wise linear coupling as well as balancing of loss-gain
between the $(2i-1)^{\textrm{th}}$ and $(2i)^{\textrm{th}}$ components of $\Psi$, where $i=1, 2, \dots m$.
It should be noted that the vector NLSE is simply $m$-copies of the system with two components
for vanishing  nonlinear interaction, i.e. $\delta=0$.  The nonlinear term $\left ( \Psi^{\dagger} M_{2m} \Psi
\right )^2 $ allows each component $\Psi_i$ to interact with all other components, including self-interaction.
We choose $\delta \neq 0$ and appropriate $M_{2m}$ such that the vector NLSE can not be expressed as $m$-copies
of the two-component system.

The pseudo-unitary operator $U_{2m}$ which connects $2m$-component
fields $\Psi$ and $\Phi$  via the relation $\Psi=U_{2m} \Phi$ is determined as
$U_{2m} := e^{-i A_{2m} t} = I_{m} \otimes U$, where $U$ is given by Eq. (\ref{uexp}).
The condition for periodic time-dependence of the matrix $U_{2m}$ is the same as that of $U$, i.e. ${\vert \beta \vert}^2 > \Gamma^2
$. We consider an autonomous NLSE (\ref{m-nlse}) with  $G_{2m}=I_m \otimes G$
and a non-autonomous NLSE (\ref{nlse}) with time-dependent $M_{2m}(t)$,
\bea
M_{2m}&:=&U_{2m}^{\dagger}(-t) G_{2m} U_{2m}(-t)\nonumber \\
& = & I_m \otimes M(t),
\eea
\noindent where $M(t)$ is given by Eq. (\ref{mexp}). Note that $M_{2m}$ has the form given in Eq. (\ref{expand-m}) and following
the general method prescribed in this article, the non-autonomous NLSE (\ref{nlse}) is mapped to autonomous NLSE (\ref{m-nlse})
by the non-unitary transformation $\Psi=U_{2m} \Phi$. The matrix $A_{2m}$ is $M_{2m}$-pseudo-hermitian, whenever the condition
(\ref{condi}) is satisfied, and in this limit $G_{2m}=M_{2m}$ leading to the result that both the Eqs. (\ref{nlse}) and
(\ref{m-nlse}) are autonomous. It is worth recalling that pseudo-hermitian operators play an important role in our understanding
of non-hermitian quantum systems admitting entirely real spectra and unitary time-evolution. The appearance of pseudo-hermitian
matrices in the context of classical system with BLG is an interesting coincidence.

The exact solution of Eq. (\ref{nlse}) may be constructed via the non-unitary transformation provided Eq. (\ref{m-nlse}) with
$G_{2m}$ is solvable. The NLSE (\ref{m-nlse}) satisfied by $\Phi_{2m}$ and $G_{2m}$ can be brought to the canonical form of
integrable Manakov-Zakharov-Schulman system by a unitary transformation followed by an appropriate scaling of the $2m$ components
of $\Phi$. In particular, the hermitian matrix $G_{2m}$ is diagonalizable by a unitary transformation 
$G_d=V^{\dagger} G_{2m} V$, where the diagonal matrix $G_d$  and the unitary matrix $V$ are given by,
\bea
&& G_d = I_m \otimes \bp \lambda_1 && 0\\ 0 && \lambda_2 \ep, \ \
V=e^{i \frac{\xi_1}{2} I_m \otimes \sigma_3} e^{i \frac{\xi_2}{2} I_m \otimes \sigma_1},\nonumber \\
&& \lambda_{i} = \alpha_0 +(-1)^{i+1} \sqrt{\alpha_1^2+\alpha_2^2+\alpha_3^2}, \ i=1, 2\nonumber \\
&&
\xi_1 \equiv \tan^{-1}\left (\frac{\alpha_1}{\alpha_2} \right ), \ \xi_2= \tan^{-1} \left ( \frac{\sqrt{\alpha_1^2+
\alpha_2^2}}{\alpha_3} \right ).
\eea
\noindent The matrix $G_{2m}$ is block-diagonal and the unitary transformation describes a $SU(2)$ rotation for each block
in terms of the $SU(2)$ generators $\frac{1}{2}(\sigma_1,\sigma_2,\sigma_3)$ \textemdash a rotation by an angle $\xi_1$ around
$\sigma_3$ followed by a rotation around $\sigma_1$ by an angle $\xi_2$. The $2m$ eigenvalues of $G_{2m}$ are $\lambda_{1,2}$
and both $\lambda_1$ and $\lambda_2$ are $m-$fold degenerate. In general, the eigenvalues $\lambda_{i}$ can take positive as well
as negative values and may be expressed as $\lambda_{i}=\sgn(\lambda_i) {\vert \lambda_{i} \vert}$. However, the negative
values of $\lambda_i$ are not allowed for the case of $M$-pseudo-hermitian $A$ for which $M$ becomes time-independent and $M=G$
due to the condition (\ref{condi}). This is because we demand entirely real eigenvalues of $A$ in order to avoid any decaying
and/or growing modes of $\Psi$ via $U=e^{-iAt}$. The eigenvalues of the matrix $M$ are required to be semi-positive definite in
order to have an entirely real eigenvalues of the pseudo-hermitian matrix $A$. The condition $\alpha_0 > {\vert \alpha \vert} 
$ ensures that the eigenvalues $\lambda_{i}$ are positive-definite and the matrix
$M=G$ is non-singular. We take only positive values of $\lambda_i$, whenever both the Eqs. (\ref{nlse}) and (\ref{m-nlse}) are
autonomous and $M=G$, otherwise allow positive as well negative values of $\lambda_i$.

A scale transformation may be used to transform $G_d$ to the diagonal form $\eta = S^{-1} G_d S^{-1}$ with eigenvalues
$\pm 1$:
\bea
&& \eta = I_m \otimes \bp \sgn(\lambda_1) && 0\\ 0 && \sgn(\lambda_2) \ep,\nonumber \\
&& S= I_m \otimes \bp \sqrt{\vert{\lambda_1}\vert} && 0\\0 && \sqrt{\vert{\lambda_2}\vert} \ep 
\eea
\noindent The NLSE (\ref{m-nlse}) is transformed to a canonical form of integrable Manakov-Zakharov-Schulman system
in terms of the field $\tilde{\Phi}=S^{-1} V \Phi$ with $2m$ components,
\bea
i \frac{\partial \tilde{\Phi}}{\partial t} =  - \frac{\partial^2 \tilde{\Phi}}{\partial x^2} -\delta
\left ( \tilde{\Phi}^{\dagger} \eta \tilde{\Phi} \right ) \tilde{\Phi}.
\label{c-nlse}
\eea
\noindent There are three distinct regions in the space of parameters:\\
(i) $\alpha_0 > {\vert \alpha \vert}$: The matrix reduces to identity matrix, i.e. $\eta=I_{2m}$ and Eq. (\ref{c-nlse})
corresponds to the Manakov system of $N$ coupled vector NLSE in a self-focusing medium.\\
(ii) $\alpha_0 < 0, {\vert \alpha_0 \vert} > {\vert \alpha \vert}$: The matrix $\eta=-I_{2m}$ and Eq. (\ref{c-nlse})
corresponds to the Manakov system of $N$ coupled vector NLSE in a self-defocusing medium.\\
(iii) $-{\vert \alpha \vert}< \alpha_0 < {\vert \alpha \vert}$: The matrix $\eta$ corresponds to the mixed case with
equal number of eigenvalues $1$ and $-1$. This corresponds to generalized Zakharov-Schulman system.\\
There are various systematic procedures\cite{kay,nogami,park} to find exact solutions of Eq. (\ref{c-nlse}) for all three cases discussed above
and many analytical solutions have been discussed in the literature. The exact solutions of Eq. (\ref{c-nlse}) may
be used to construct exact solutions of Eq. (\ref{nlse}) by using the relation,
\bea
\Psi= U_{2m} V S^{-1} \tilde{\Phi}.
\eea
\noindent We have already presented exact analytical expressions of $\Psi$ for $N=2$. Exact solutions of NLSE with BLG
along with the time-dependence of different observables for $N >2$ will be presented in a separate publication.

To conclude, we have presented a generic method to remove loss-gain and LC terms from a
vector NLSE by a time-dependent non-unitary transformation which imparts time-dependence to the
nonlinear term. Further, if the generator of the transformation is pseudo-unitary, the non-linear
term remains unchanged even though the loss-gain and LC terms are completely removed. 
The method is applicable to a class of vector NLSE with cubic nonlinearity that is subjected to BLG
and LC, and useful to construct solvable models. We have constructed an exactly solvable
two-component NLSE with BLG and LC that exhibits power-oscillation.
Exactly solvable models of NLSE with more than two components and subjected to BLG
have also been constructed.

The inclusion of more generalized cubic nonlinear interaction in Eq. (\ref{nlse}) may be achieved
by replacing $\Psi^{\dagger}M\Psi$ with $K$, where $K$ is a $N\times N$ hermitian matrix with
elements $[K]_{ij}=\Psi^{\dagger} H_{ij} \Psi$ and $H_{ij}$ are $N^2$ constant hermitian matrices of
dimension $N\times N$. The Eq. (\ref{nlse}) may or may not admit a Hamiltonian for a generic $K$.
The matrix $K$ can be re-expressed as $[K]_{ij}=\Phi^{\dagger}
\left (U^{\dagger} H_{ij} U \right )\Phi$. The BLG and LC terms are completely removed by the non-unitary
transformation at the cost of imparting time-dependence to the nonlinear term.
If $U$ is pseudo-unitary with respect to each matrix $H_{ij}$, then $[{K}]_{ij}=
\Phi^{\dagger} H_{ij} \Phi$ remains form invariant. The BLG and LC are removed by
the pseudo-unitary transformation and the nonlinear term $K \Psi$ is changed to
$U^{-1} K U \Phi$ with  $[{K}]_{ij}= \Phi^{\dagger} H_{ij} \Phi$. The nonlinear term
is time-independent and $K \Psi \rightarrow K \Phi$ only if $[K,U]=0 \Rightarrow [K,A]=0$.

The results can be trivially generalized to higher spatial dimensions and/or by including
a space-time dependent inhomogenous term $\lambda_0 V(x,t) \Psi$. The time modulated gain-loss strength
and LC can be implemented by replacing the non-hermitian matrix $A$ with
$\tilde{A}(t)=\int dt A(t)$ in the definition of $U(t)$ and for all subsequent steps.
Investigations along these directions could be carried out by using the method
prescribed in this article to explore a wide variety of physically interesting models with BLG.

\begin{acknowledgments}
This work is supported by a grant ({\bf SERB Ref. No. MTR/2018/001036})
from the Science \& Engineering Research Board(SERB), Department of Science
\& Technology, Govt. of India under the {\bf MATRICS} scheme.
\end{acknowledgments}

\end{document}